\documentclass[numberedappendix]{emulateapj}
\usepackage{natbib}
\usepackage{epsfig}
\usepackage{amsmath}
\usepackage{color}

\shorttitle {On the Dearth of Close-In Planets around Fast Rotators}
\shortauthors{Teitler and K\"{o}nigl}

\begin{document}
\title{Why is there a Dearth of Close-In Planets around Fast-Rotating stars?}

\author{Seth Teitler\altaffilmark{1} and Arieh K\"{o}nigl\altaffilmark{2}}

\altaffiltext{1}{National Academic Quiz Tournaments, LLC, 11521 W 69th Street, Shawnee, KS 66203, USA; satelite@gmail.com}
\altaffiltext{2}{Department of Astronomy \& Astrophysics and The Enrico Fermi Institute, The University of Chicago, Chicago, IL 60637, USA; akonigl@uchicago.edu}
\begin{abstract} 
We propose that the reported dearth of {\it Kepler}\/ Objects of Interest (KOIs) with orbital periods $P_{\rm orb} \lesssim 2-3\;$days around stars with rotation periods $P_{\rm rot} \lesssim 5-10\;$days can be attributed to tidal ingestion of close-in planets by their host stars. We show that the planet distribution in this region of the $\log{P_{\rm orb}}-\log{P_{\rm rot}}$ plane is qualitatively reproduced with a model that incorporates tidal interaction and magnetic braking as well as the dependence on the stellar core--envelope coupling timescale. We demonstrate the consistency of this scenario with the inferred break in the $P_{\rm orb}$ distribution of close-in KOIs and point out a potentially testable prediction of this interpretation.
\end{abstract}

\keywords{planets and satellites: dynamical evolution and stability -- planet--star interactions -- stars: rotation}

\section{Introduction}
\label{sec:intro}

The {\it Kepler}\/ space telescope has already led to tremendous advances in the detection and characterization of extrasolar planetary systems and in the study of the properties of their host stars. Until recently, these two aspects of the observed systems were considered separately, but the accumulation of data on both planets and stars is starting to make it possible to carry out statistical investigations of their joint characteristics. Such investigations can potentially shed new light on how these systems form and evolve, and, in particular, on how planets interact with their host stars.

In a recent study, \citet[][hereafter MMA13]{McQuillanEtal13} derived the rotation period $P_{\rm rot}$  for 737 host stars of {\it Kepler}\/ Objects of Interest (KOIs) by applying an autocorrelation-function technique to the analysis of the observed star spot-induced photometric modulations. After plotting these periods against the orbital period $P_{\rm orb}$ of the innermost planet in each system, they noticed a clear dearth of close-in planets ($P_{\rm orb}\la 2-3\;$days) around rapidly rotating stars ($P_{\rm rot}\la 5-10\;$days). They fitted a line with a slope of $-0.69$ to the lower edge of the observed KOI distribution in the region of the $\log{P_{\rm orb}}-\log{P_{\rm rot}}$ plane bounded by $P_{\rm orb}\le 10\;$days and $P_{\rm rot} \ge 3 \;$days. They also pointed to the presence of several objects below the lower edge that exhibit near-synchronous rotation ($P_{\rm rot} \approx P_{\rm orb}$). These results were confirmed by \citet{WalkowiczBasri13}, who deduced the rotation periods of $\sim 950$ KOI hosts using the Fourier-based periodogram method. The latter authors found a compelling dearth of planets with $P_{\rm orb}$ and $P_{\rm rot}$ periods $< 6\;$days, and also noticed a concentration of planets with $P_{\rm rot} \approx P_{\rm orb}$. They furthermore pointed to the presence of a few systems with $P_{\rm rot} \approx 2\, P_{\rm orb}$, and noted that all the planets  along these two loci have radii $R_{\rm p}>6\, R_\earth$.

In this paper we propose that the short-period void in the $\log{P_{\rm orb}}-\log{P_{\rm rot}}$ plane of KOIs can be attributed to the tidal interaction between close-in planets and their (typically Sun-like) host stars, which, over the lifetimes of the observed systems, results in the spiraling-in of the nearest planets and the deposition of their orbital angular momenta in the host's envelope. Since the angular momentum of a planet of a given mass that moves on a Keplerian orbit scales as $P_{\rm orb}^{1/3}$ and its inspiral onset time is $\propto P_{\rm orb}^{13/3}$ (see Section~\ref{sec:formulate}), progressively lower values of $P_{\rm rot}$ correspond to the tidal ``ingestion'' of planets with progressively larger values of $P_{\rm orb}$. This naturally results in an inverse correlation between $P_{\rm rot}$ and the orbital period of the surviving closest-in planet. Using the inferred parameters of KOIs in a tidal interaction model, we demonstrate that this picture can indeed account for the lower edge of the empirical planet distribution.

Our model also incorporates two physical processes that affect the stellar envelope and act to counter the spinup effect of planet ingestion. The first is core--envelope coupling, through which the envelope shares its angular momentum with the rest of the star. The second is magnetic braking, the process invoked to account for the apparent rotation--age correlation underlying the gyrochronology age-determination method for solar-type stars \citep[e.g.,][]{MeibomEtal11a,MeibomEtal11b}. The braking time is typically $<10^9\,$yr when $P_{\rm rot}<10\;$days, which explains our finding that the systems at the lower edge of the close-in planet distribution are generally among the youngest in the sample (see Section~\ref{sec:results}). The relative efficiency of this process would, however, inhibit the formation of synchronous systems, and we suggest (in line with previous work) that the observed $P_{\rm rot} \approx P_{\rm orb}$ locus likely corresponds to systems in which both magnetic braking and core--envelope coupling are weak.

The possibility that the spin evolution of a star can be affected by the tidally induced ingestion of close-in planets was already considered previously in the literature. In particular, \citet{JacksonEtal09} suggested that this process could account for the observed orbital distribution of close-in planets and pointed out the dependence of the results on the initial $P_{\rm orb}$ distribution as well as on the systems' age distribution, \citet{Pont09} emphasized the dependence of the tidal interaction on the planet's mass, and \citet{BolmontEtal12} highlighted the potential implications of stellar spinups of this type to the reliability of the gyrochronology method. In this contribution we focus on the application of this idea to the distribution of low-period KOIs in the $\log{P_{\rm orb}}-\log{P_{\rm rot}}$ plane.

\section{Formulation}
\label{sec:formulate}

We construct a simplified computational scheme that aims to capture the main physical ingredients of the proposed stellar spinup mechanism and its dependence on relevant parameters. We consider planets that move on circular orbits in the star's equatorial plane and neglect the contributions of planetary spin to the total angular momentum, of tidal dissipation within the planets, and of possible planet--planet interactions, which should be of secondary importance to the modeled process. The assumption of a circular orbit is motivated by the expectation that the timescale for tidal capture is determined by the initial value of the planet's semimajor axis $a$ rather than by its initial eccentricity \citep{JacksonEtal09}, and it is supported by the inference from observations that orbit circularization may be faster than orbital decay \citep[e.g.,][]{MatsumuraEtal10}. We approximate the star as consisting of an envelope with moment of inertia $I_{\rm e}$ and rotation period $P_{\rm rot}$, and a core with a moment of inertia $I_{\rm core} = I_*-I_{\rm e}$ and rotation period $P_{\rm core}$, neglecting the possible changes of $I_{\rm e}$ and $I_{\rm core}$ with time. 

The basic evolution equations are
\begin{equation}
\label{eq:Porb}
\frac{dP_{\rm orb}}{dt} = -\frac{27\pi}{Q^\prime_*}\frac{M_{\rm p}}{M_*}\frac{R_*^5}{a^5} \left (1 - \frac{P_{\rm orb}}{P_{\rm rot}} \right )
\end{equation}
\begin{eqnarray}
\label{eq:Prot}
\frac{dP_{\rm rot}}{dt} =  &-& \frac{9\pi}{Q^\prime_*} \frac{M_{\rm p}}{M_*+M_{\rm p}}\frac{M_{\rm p}R_*^2}{I_{\rm e}}\frac{R_*^3}{a^3}\frac{P_{\rm rot}^2}{P_{\rm orb}^2}\left(1- \frac{P_{\rm orb}}{P_{\rm rot}}\right )\nonumber \\
&-& \frac{P_{\rm rot}^2 N_{\rm mag}}{2\pi I_{\rm e}} + \left (1- \frac{I_{\rm e}}{I_*}\right )\frac{P_{\rm rot}}{\tau_{\rm c}}\left(1- \frac{P_{\rm rot}}{P_{\rm core}}\right )
\end{eqnarray}
\begin{equation}
\label{eq:Pcore}
\frac{dP_{\rm core}}{dt}=\frac{I_{\rm e}}{I_*}\frac{P_{\rm core}}{\tau_{\rm c}}\left(1-\frac{P_{\rm core}}{P_{\rm rot}}\right )\ ,\end{equation}
where $M_*$ and $M_{\rm p}$ are the stellar and planetary mass, respectively, $R_*$ is the stellar radius, $N_{\rm mag}$ is the magnetic braking torque, and $\tau_{\rm c}$ is the core--envelope coupling time. The term on the right-hand side of Equation~(\ref{eq:Porb}) and the first term on the right-hand side of Equation~(\ref{eq:Prot}) represent the effect of tidal friction in the star on $P_{\rm orb}$ and $P_{\rm rot}$, respectively. The tidal interaction is modeled assuming quasi-hydrostatic equilibrium tides \citep[e.g.][]{Hut81}, and its strength is parametrized by the normalized tidal quality factor $Q^\prime_*\equiv 1.5\,Q/k_2$ (where $k_2$ is the second-order potential Love number). For simplicity, we neglect the possible contribution of dynamical tides associated with the excitation of stellar oscillations \citep[e.g.][]{WitteSavonije02,OgilvieLin07}. As is evident from the form of the aforementioned terms, the effect of the tidal interaction depends on the relative magnitudes of $P_{\rm orb}$ and $P_{\rm rot}$. In our model setup, each planet initially satisfies $P_{\rm orb}<P_{\rm rot}$ and therefore moves inward while acting to spin up the star. However, as $P_{\rm rot}$ goes down, it can happen that the next closest planet finds itself outside the corotation radius (where $P_{\rm orb}=P_{\rm rot}$) and starts to move out (with tidal friction now acting to increase $P_{\rm rot}$). But even in this case, magnetic braking soon pushes the corotation radius back beyond the planet, resulting in the resumption of the planet's inward motion (with the tide again acting to spin up the star). The second term on the right-hand side of Equation~(\ref{eq:Prot}) represents the effect of magnetic braking, whereas the last term on the right-hand side of this equation and the term on the right-hand side of Equation~(\ref{eq:Pcore}) represent the effect of core--envelope coupling. We follow standard treatments of angular momentum transport in solar-type stars \citep[e.g.,][]{DenissenkovEtal10} in the way we model these terms. In particular, we express the magnetic braking torque in the form
\begin{equation}
\label{eq:Nmag}
N_{\rm mag} = - K_{\rm w} (R_*/R_\sun)^{0.5}(M_*/M_\sun)^{-0.5} {\rm min} (\Omega_{\rm e}\Omega_{\rm sat}^2, \Omega_{\rm e}^3)\,,
\end{equation}
where $K_{\rm w}$ is a calibration constant, $\Omega_{\rm e} = 2\pi/P_{\rm rot}$ is the envelope's angular velocity, and $\Omega_{\rm sat}$ 
accounts for the inferred saturation of the surface magnetic field at high stellar rotation rates. We treat the quantities $\tau_{\rm c}$, $K_{\rm w}$, and $\Omega_{\rm sat}$ as model parameters.

We derive the KOI distribution in the $\log{P_{\rm orb}}-\log{P_{\rm rot}}$ plane by performing Monte Carlo experiments that employ the parametrized planetary distribution function $\partial f / \partial {\ln{R_{\rm p}}}\partial{\ln{P_{\rm orb}}}  \propto R_{\rm p}^\alpha P_{\rm orb}^\beta$ that was constructed by \citet{Youdin11} for $P_{\rm orb}$ in the range $0.5-50\;$days and $R_{\rm p}$ in the range $2-20\, R_\earth$ using data for 372 KOIs and the discovery efficiency estimates of \citet{HowardEtal12}. Although the adopted distribution neglects the contribution of Earth-size KOIs, the effect of the latter on the stellar spinup process should be relatively small given that the timescale for the onset of a planet's inspiral is $\propto M_{\rm p}^{-1}$ (Equation~(\ref{eq:Porb})) and that the angular momentum deposited at ingestion is $\propto M_{\rm p}$ (Equation~(\ref{eq:Prot})). The observed period distribution exhibits a sharp decrease in planet counts for $P_{\rm orb} \lesssim 3\;$days, which motivated \citet{Youdin11} to divide his sample into ``slow'' and ``fast'' groups (separated somewhat arbitrarily at $P_{\rm orb}=7\;$days on account of binning considerations). Since at least  some of the increase in the exponent $\beta$ for ``fast'' planets could be due to the tidal ingestion process that we study (see \citealt{JacksonEtal09}), we draw our candidate planets only from Youdin's ``slow'' distribution (for which $\beta \approx 0.54$), which we extrapolate all the way down to the 
Roche limit $a_{\rm R} = (3R_{\rm p}/2)[M_{\rm p}/3(M_*+M_{\rm p})]^{-1/3}$ \citep[e.g.,][]{MatsumuraEtal10}.\footnote{We assume for simplicity that a planet reaching $a_{\rm R}$ is destroyed and that its angular momentum is incorporated into the stellar envelope. In reality, the situation could be more complicated in certain cases \citep[e.g.,][]{TrillingEtal08,MetzgerEtal12}.} We check this approach a posteriori by examining the shape of our predicted final $P_{\rm orb}$ distribution (see Section~\ref{sec:discuss}). 

\citet{Youdin11} noticed that the planet size distribution also changes for low values of $P_{\rm orb}$ and suggested that this behavior can, at least in part, be attributed to the effect of photoevaporation. Since this process is likely to operate on timescales $\lesssim 10^8\,$yr \citep[e.g.,][]{OwenWu13,LopezFortney13}, which are short compared to the typical tidal ingestion time, we adopt the size exponents inferred by \citet{Youdin11} for the ``slow'' and ``fast'' sides of $P_{\rm orb}=7\;$days ($\alpha \approx-2.31$ and $-1.09$, respectively). To convert from planetary size to mass, we use the scaling $M_{\rm p} \approx M_\earth (R_{\rm p}/R_\earth)^2$ (see \citealt{Youdin11} and \citealt{LissauerEtal11}). With this scaling, we find that the final average mass per planetary system in our fiducial model is $\sim 34\, M_\earth$. This value is consistent (within a factor of $\sim 2$) with the average mass of $M_{\rm p} \ge M_\earth$ planets with $P_{\rm orb}\le10\;$days around planet-bearing stars that is inferred from radial-velocity surveys \citep[e.g.,][]{MayorEtal11,WrightEtal12}, which suggests that most of the mass associated with close-in planets is accounted for in our scheme. 

Given that, in the proposed scenario, a low-$P_{\rm rot}$ system must have initially contained at least two planets, we simplify our treatment by only considering systems that at an early time contain 2 (or possibly 3) planets with $P_{\rm orb} \le 10\;$days and $a>a_{\rm R}$. We first determine each planet's initial values of $P_{\rm orb}$ and $R_{\rm p}$ using independent drawings from the adopted distribution function (a procedure consistent with the fact that this distribution was constructed by treating planet occurrence as a Poisson process; see discussion in \citealt{Youdin11}). We pick the initial value of $P_{\rm rot}$ by adopting $\tau_0=1\,$Gyr as the starting time and assuming that the stellar rotation periods at that age are distributed uniformly in the interval $10-12\;$days \citep[see][]{MeibomEtal11a}; we also assume that the rotation rate for each star is initially the same for the envelope and the core. We then pick a value $\tau>\tau_0$ for the age of the modeled system from the inferred distribution (which peaks at $\sim 1.5-2\,$Gyr) of the \citet{WalkowiczBasri13} KOI sample. This  age distribution was deduced using the gyrochronology method, which, as we already noted, can be corrupted by the tidal ingestion process. However, we assume that this effect does not lead to major deviations from the correct distribution (and return in Section~\ref{sec:results} to check on the validity of this assumption). Our choice of $\tau_0$ is motivated by the fact (emphasized in \citealt{WalkowiczBasri13}) that gyrochronology age determinations for stars with $P_{\rm rot} < 10\;$days, which correspond to ages $\lesssim 1\,$Gyr for Sun-like hosts, are not well constrained \citep[e.g.,][]{MeibomEtal11b,GalletBouvier13}. One should, however, keep in mind that, in reality, the tidal interaction may well start when a system is still $<1\,$Gyr old. In the final step, we follow the planets' evolution over the time interval $\tau-\tau_0$ using Equations~(\ref{eq:Porb})--(\ref{eq:Pcore}).

\section{Results}
\label{sec:results}

\begin{figure*}
\centering
\includegraphics[totalheight=0.75\textwidth, angle=0,]{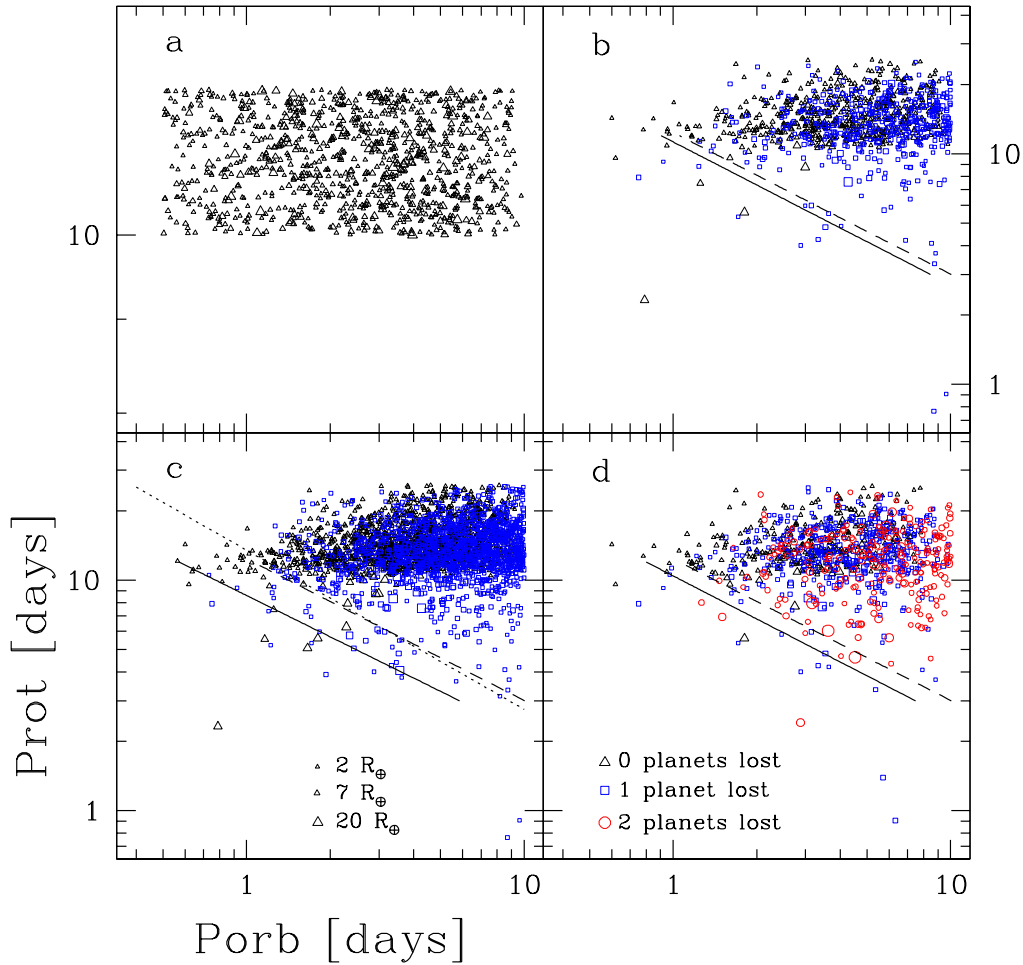}
\caption{Properties of fiducial model (see Table~\ref{tab1}). (a) Initial distribution for 1000-system sample. (b) 1000-system sample for Model~1 (2-planet drawings). (c) 3000-system sample for Model~1 (which includes the 2000~systems shown in Figure~\ref{fig3}(a) and hence also the 1000~systems shown in panel~(b)). (d) 1000-system sample for Model~2 (3-planet drawings). Each symbol represents a single system, with its size corresponding to that of the innermost surviving planet, and with its shape and color corresponding to the number of ingested planets. The solid lines represent our fit to the lower edge of the plotted distribution in the parameter-space region \{$P_{\rm orb}\le 10\;$days, $3\;{\rm days} \le P_{\rm rot} \le 12\;$days\}, whereas the dashed lines show our corresponding fit to the data in MMA13. The dotted line in panel~(c) reproduces the fit that \citet{McQuillanEtal13} obtained for their data without imposing an upper bound on $P_{\rm rot}$ and using a different edge-finding method. The parameters of the linear fits are given in Table~\ref{tab2}.}
\label{fig1}
\end{figure*}

\begin{figure*}
\centering
\includegraphics[totalheight=0.75\textwidth, angle=0]{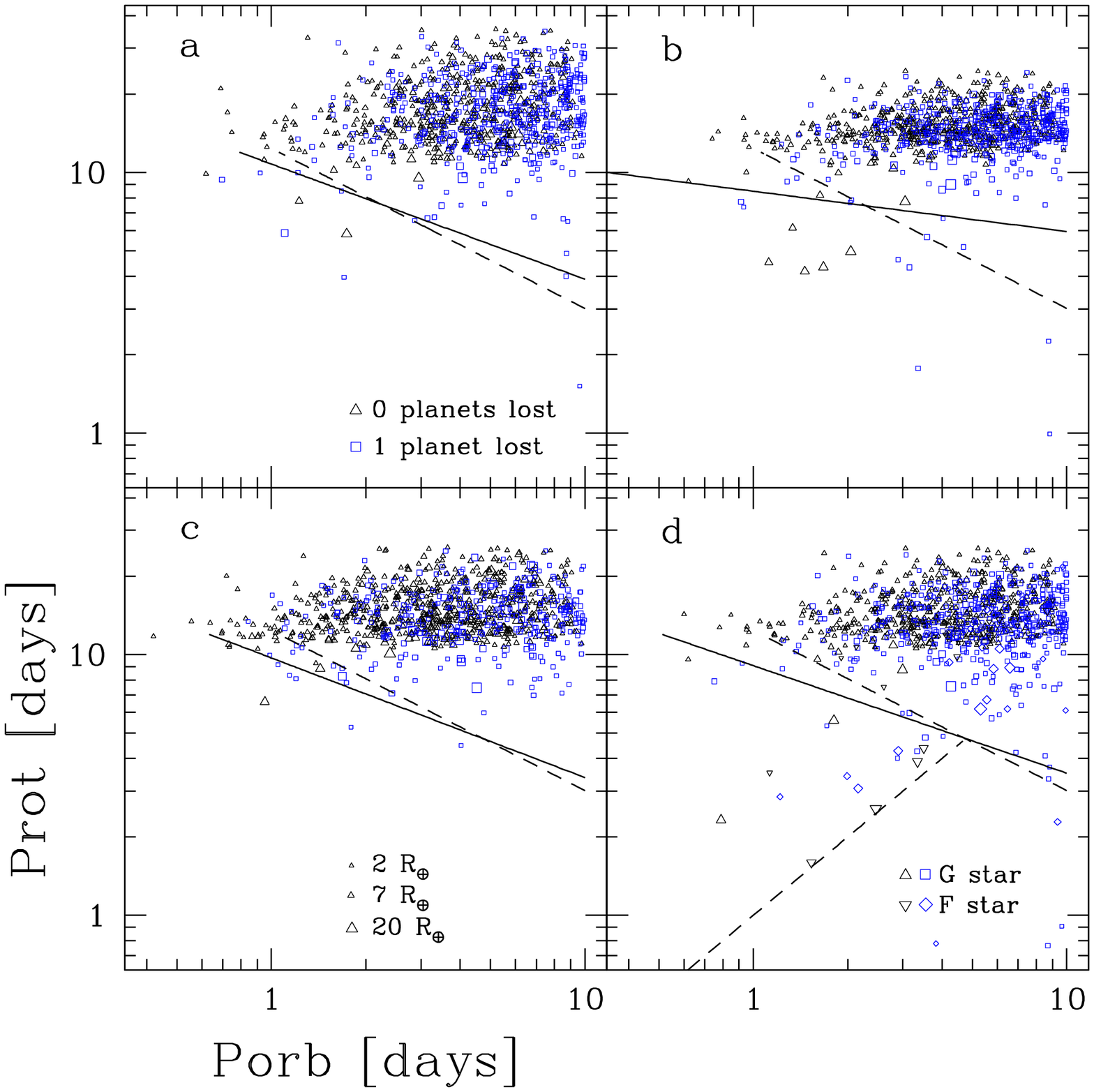}
\caption{Parameter dependence of final distribution (see Table~\ref{tab1}; all 1000-system samples). (a) Model~3. (b) Model~4. (c) Model~5. (d) Hybrid model: 95\% G~dwarfs (Model~1), 5\% late-F~dwarfs (Model~6). Notation is the same as in Figure~\ref{fig1}. In panel~(d), the added dashed line represents the synchronicity locus ($P_{\rm rot}=P_{\rm orb}$), and systems with F-type hosts are shown by symbols with rotated orientations.}
\label{fig2}
\end{figure*}

The physical parameters of our modeled systems are given in Table~\ref{tab1}. The values of $\tau_{\rm c}$ that we use ($10^7$ and $10^8\,$yr) roughly bracket those inferred for solar-mass stars \citep[e.g.,][]{DenissenkovEtal10,GalletBouvier13}. Similarly, our two adopted $\{K_{\rm w},\, \Omega_{\rm sat}\}$ combinations correspond to comparatively weak (Model~1) and strong (Model~3) magnetic braking models for such stars. We use $I_* = 5.9\times 10^{-2}\, M_\sun\,R_\sun^2$ and $I_{\rm e}=6.6\times 10^{-3}\, M_\sun\,R_\sun^2$ for Sun-like hosts \citep[see][]{WinnEtal10}, and reduce $I_{\rm e}$ by a factor of 10 for late F-type stellar hosts (Model~6; see \citealt{BarkerOgilvie09}). For each parameter combination that we consider, we perform a Monte Carlo experiment involving $10^6$ systems drawn from the chosen initial distribution. For the 2-planet experiments, we construct two such pools, each consisting of $10^6$ systems with specified stellar rotation and planetary masses and locations --- one corresponding to G-type hosts and the other to F-type hosts. The F-host pool is used in constructing the F-star admixture considered in Figure~\ref{fig2}(d), whereas all the other systems that we model (including the rest of the ones included in Figure~\ref{fig2}(d)) are drawn from the G-host pool. 
In the case of the 3-planet experiment considered in Figure~\ref{fig1}(d), we add a third planet to each of the $10^6$ systems in the aforementioned G-host pool. After evolving these systems, we find that a fraction (ranging from $\sim 4\%$ to $\sim 12\%$) end up with no surviving planets and are therefore excluded from further consideration. The remaining ones are used as a parent population for the samples that we analyze, which typically consist of only $1000$ planetary systems to facilitate comparison with existing observational data. For the 1000-system samples that we present in the figures, we pick the first $1000$ systems from the respective parent populations (i.e., the first $1000$ systems drawn from the original pool that have at least one surviving planet at the end of the evolution for the given parameter combination). These samples therefore have very similar initial conditions --- they typically differ only at the few-percent level on account of differences in the parameter-dependent number of systems with no surviving planets at the end of the evolution. Similar considerations govern the construction of the 2000-system and 3000-system samples that we display.

MMA13 fitted a line to the lower edge of the observed KOI distribution in the $\log{P_{\rm orb}}-\log{P_{\rm rot}}$ plane by minimizing a function that incorporated the perpendicular distance between the line and data points, and used different weights for points above and beneath the line so that $95\%$ of the points obeying their selection criteria ended up lying above the line. However, when this procedure is applied (with the same weights) to our model distributions, it typically produces a line that 
deviates significantly from the apparent lower edge of the distribution. This is evidently a consequence of the fact that the overall properties of the respective distributions are different; in particular, as discussed below, our derived distributions are noticeably sparser in comparison with the MMA13 data in the parameter-space region of interest. In view of this fact, we implement a procedure that probes only the points in the vicinity of the lower edge of the distribution. Specifically, we divide the $P_{\rm rot}$ range between~3 and~$12\;$days into 9 equal intervals, locate the 3 data points with the smallest values of $P_{\rm orb}\le10\;$days in each of them and form their averages, and then fit a least-squares regression line to the logarithms of these averages.
In cases where fewer than 3 points can be located within a given interval, the available points are added to the tally of the adjacent, higher-$P_{\rm rot}$ interval, and the averaging is carried out over the total number of selected points in the two intervals. The linear fit has the form $\log{P_{\rm rot}} = m\log{P_{\rm orb}}+c$, and we determine the uncertainties in the fitting parameters $m$ and $c$ by calculating their means and standard deviations (SDs) based on fits to $10^3$ randomly generated data sets, each with the prescribed number of systems (1000, 2000, or 3000), drawn from the parent populations produced in our Monte Carlo experiments. For our model fit to the MMA13 data, we estimate uncertainties using the method employed by MMA13 (generate fits using $10^3$ random drawings of $80\%$ of the points in the selected parameter-space region). The estimates of $m$ and $c$ obtained in this way are collected in Table~\ref{tab2}. For comparison, the first entry in the table lists the corresponding values derived by applying the MMA13 fitting procedure to the observed distribution and using their prescribed parameter ranges ($P_{\rm orb}\le 10\;$days, $P_{\rm rot}\ge 3\;$days). It is seen that, while our method yields a somewhat flatter fit to these data, the 1-SD zones of the respective $m$ and $c$ values overlap.\footnote{Note also that, for both of these fits, the value of $m$ lies within 1~SD from $<m>$.}

\begin{figure}
\includegraphics[width=0.5\textwidth,height=0.5\textwidth,angle=0]{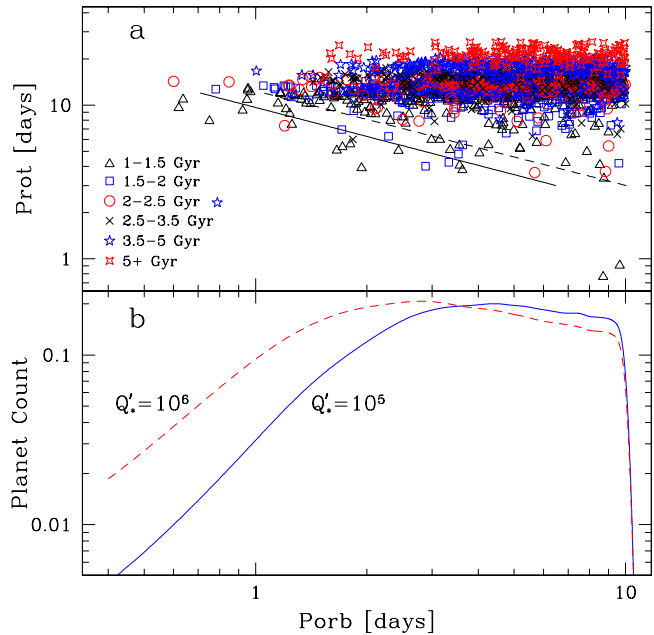}
\caption{(a) Age distribution for a sample of 2000~systems in Model~1 (which includes the 1000~systems shown in Figure~\ref{fig1}(b)), grouped according to the indicated symbol shapes and colors. The solid and dashed  lines have the same meanings as in Figure~\ref{fig1}. (b) Final planet-count distribution for Model~1 (solid blue curve); the average number of planets with $P_{\rm orb}\le 10\;$days is 1.5 per star. The dashed red curve shows the corresponding distribution for Model~5.} 
\label{fig3}
\end{figure}

Following a series of experiments, in which we compared the derived values of $m$ and $c$ with those obtained from our fit to the MMA13 data, we identified the parameters defining Models~1 and~2 as producing the best match to the lower edge of the observed KOI distribution. Figure~\ref{fig1} illustrates how the physical processes incorporated into our model transform the adopted narrow initial $P_{\rm rot}$ distribution (panel~(a)) into one that, at least qualitatively, resembles the data in Figure~2 of MMA13.\footnote{We reiterate that the results presented in the figures correspond in each case to the first data set with the prescribed number of systems that we selected from the initial pool of $10^6$ systems. One can gauge how representative each of these sets is by comparing the lower-edge linear fit parameters for the plotted distribution with the mean values deduced on the basis of all the $10^3$ data sets that we generated for the given case through random drawings from the parent population (see Table~\ref{tab2}). This comparison reveals that the slopes of the linear fits for the displayed model realizations are typically steeper than the corresponding mean slopes (as is also the case for our fits to the MMA13 data), but that in most instances the line fitting parameters lie within 1~SD from their respective means (or nearly so).} In particular, the lower edge identified by MMA13 is already clearly discernible in a sample of $1000$ systems derived from a 2-planet drawing experiment (panel~(b)), and its outline for $P_{\rm rot}\lesssim 5\;$days is determined by systems (indicated by blue squares) that have already lost one planet to the star, as expected in the planet-ingestion scenario.\footnote{46.5\% of the total systems shown in Figure~\ref{fig1}(b) have already lost a planet; in Figure~\ref{fig1}(d), 43.6\% lost one and 23.5\% lost two.} Also as expected (given that the edge of the distribution is defined by relatively rare events), the trend becomes visually clearer when the sample is increased to 3000 systems (panel~(c); the distribution for 2000 systems is shown in Figure~\ref{fig3}(a)), and when the experiment involves 3-planet drawings (which increases the probability of locating planets close to the star at time $\tau_0$; panel~(d)). In the larger sample shown in panel~(c), one also finds more low-$P_{\rm rot}$ systems that did not yet lose any planets --- these represent comparatively massive planets that are caught on their way  in toward eventual ingestion. (As we already noted in Section~\ref{sec:formulate}, the timescale for the onset of inspiral is $\propto M_{\rm p}^{-1}$.) There are, however, no systems with $P_{\rm rot}<10\;$days and $P_{\rm orb} \gtrsim 3\;$days in which all initial planets are still intact --- for planets at these distances, tidal interaction remains ineffective, and the only way these systems can correspond to a low value of $P_{\rm rot}$ is if they already lost at least one closer-in planet. We thus identify $P_{\rm orb} \approx 3\;$days as the approximate value of (or, more accurately, a lower bound on) the extent of the tidal interaction zone in this  model. This value is consistent with the apparent extent of the void in Figure~2 of MMA13. It is also worth noting that even the region $P_{\rm rot}\ge10\;$days contains a significant fraction of systems that are missing one planet (and, in the case of the 3-planet drawing experiment shown in panel~(d), also a measurable number of systems that have lost two planets) --- these correspond to spun-up stars that have already been spun back down by magnetic braking.

\begin{deluxetable}{crrlcc}
\tabletypesize{\small}
\tablewidth{0.5\textwidth}
\tablecolumns{6}
\tablecaption{Model Parameters\label{tab1}}
\tablehead{\colhead{Model} & \colhead{$\frac{\tau_{\rm c}}{{\rm Myr}}$} & \colhead{$\frac{Q^\prime_*}{10^5}$} & \colhead{$\frac{K_{\rm w}}{10^{47}\, {\rm cm}^2\, {\rm g \, s}}$} & \colhead{$\frac{\Omega_{\rm sat}}{\Omega_\sun}$} & \colhead{Planets per draw}}
\startdata
1 & 10 & 1 & \quad 1.25\tablenotemark{a} & \  8\tablenotemark{a} & 2 \\
2 & 10 & 1 & \quad 1.25 & 8 & 3 \\
3 & 10 & 1 & \quad 2.70\tablenotemark{b} & 14\tablenotemark{b} & 2 \\
4 & 100 & 1 & \quad 1.25 & 8 & 2 \\
5 & 10 & 10 & \quad1.25 & 8 & 2 \\
6 & $\infty$ & 1 & \quad 0.03 & 8 & \quad\quad 2
\tablenotetext{a}{\citet{DenissenkovEtal10}.}
\tablenotetext{b}{\citet{BouvierEtal97}.}
\enddata
\end{deluxetable}

\begin{deluxetable}{crrrc}
\tabletypesize{\small}
\tablewidth{0.5\textwidth}
\tablecolumns{6}
\tablecaption{Parameters of Linear Fits to the Lower Edge\label{tab2}}
\tablehead{\colhead{Distribution}&\colhead{$m$}&\colhead{$c$}&\colhead{$<m>\pm\;\delta m$}&\colhead{$<c>\pm\;\delta c$}}
\startdata
KOI data\tablenotemark{a} & -0.69  & 1.13 & $ -0.64 \pm 0.08$ & $1.13 \pm 0.03$ \\
KOI data\tablenotemark{b}  & -0.61& 1.09 & $-0.53 \pm 0.12$ & $1.09 \pm 0.05$\\
Figure~\ref{fig1}(b) & -0.62  & 1.05& $-0.49 \pm 0.11$ & $1.01 \pm 0.03$\\
Figure~\ref{fig1}(c) & -0.59 & 0.93& $ -0.50\pm 0.11$&$1.01 \pm 0.03$\\
Figure~\ref{fig1}(d) & -0.62 & 1.02 & $-0.47 \pm 0.10 $&$1.00 \pm 0.03$\\
Figure~\ref{fig2}(a) & -0.44 & 1.03& $-0.33 \pm 0.11$&$1.03 \pm 0.03$\\
Figure~\ref{fig2}(b) & -0.15 & 0.93 &$ -0.46\pm 0.16$&$1.02 \pm 0.04$\\
Figure~\ref{fig2}(c) & -0.46 & 0.99&$-0.35 \pm 0.08$&$0.97 \pm 0.02$\\
Figure~\ref{fig2}(d) & -0.41 & 0.96 &$-0.49 \pm 0.11$&$1.01 \pm 0.03$\\
Figure~\ref{fig3}(a) & -0.63 & 0.99 &$-0.49 \pm 0.11$&\quad\quad $1.01 \pm 0.03$ 
\tablenotetext{a}{MMA13 fitting method.}
\tablenotetext{b}{Our fitting method.}
\enddata
\end{deluxetable}


The dependence of the final distribution on the model parameters $K_{\rm w}$, $\Omega_{\rm sat}$, $\tau_{\rm c}$, and $Q^\prime_*$ is illustrated in Figure~\ref{fig2} (panels~(a)--(c)). It is seen that, as the spindown efficiency of the stellar envelope increases, either because the magnetic braking is stronger or because the core--envelope coupling time is longer, there are fewer systems with $P_{\rm rot}<10\;$days. This results in a more horizontal lower edge than in the fiducial case and therefore in a worse fit to the KOI data.\footnote{Note from Table~\ref{tab2} that the realization of Model~4, shown in Figure~\ref{fig2}(b), deviates from the other examples in having a flatter slope than the mean, with a value of $m$ that differs from $<m>$ by nearly 2~SDs. The value of $<m>$ for this case is, however, fairly close to that of the fiducial model, although $\delta m$ is relatively large.} A decrease in the tidal interaction strength (i.e., a larger $Q^\prime_*$) has a similar effect. It is, however, noteworthy that even when the model parameters are changed away from their ``optimal'' choices, the mean values $<m>$ and $<c>$ of the linear-fit parameters are consistent with each other in that their 1-SD uncertainty zones overlap (see Table~\ref{tab2}). There is also a corresponding overlap (or very nearly so in the case of Model~5 shown in Figure~\ref{fig2}(c)) with the mean edge-fitting parameters that we derived for the MMA13 data. This indicates that the general outline of this edge is a robust feature of this model, grounded in the underlying basic processes of tidal ingestion and magnetic braking.

Our finding that the proposed interpretation of the lower edge favors values of $\tau_{\rm c}$ closer to $10^7\,$yr than to $10^8\,$yr is consistent with recent inferences from the modeling of the rotational evolution of Sun-like stars \citep[e.g.,][]{GalletBouvier13}. As regards the related finding that weaker magnetic braking better accounts for the shape of the edge, we note that \citet{CohenEtal10} argued that the presence of a close-in, magnetized, giant planet could reduce the efficiency of magnetic braking in the host star. It is thus conceivable that the systems defining the lower edge of the observed distribution are characterized by low effective values of $K_{\rm w}$ and/or $\Omega_{\rm sat}$, but that many of the systems that are not near this edge are subject to stronger braking. Based on Figure~\ref{fig2}(a), this would push the upper range of the predicted distribution to higher values of $P_{\rm rot}$,  closer to the upper range ($\lesssim 50\;$days) of the data presented in Figure~2 of MMA13.

Another difference between the distributions that we derive and the one shown in Figure~2 of MMA13 is the relative sparseness of the parameter-space region $P_{\rm rot}<10\;$days in our model results. For example, the parameter range used in our linear fits ($P_{\rm orb} \le 10\;$ days; $P_{\rm rot}$ between~3 and~$12\;$days) contains 75 points in the case of the MMA13 KOI data and 220 systems in our fiducial model; out of these, the number of systems with $P_{\rm rot} < 10\;$days is 45 and 67, respectively, indicating that our model distribution is sparser in this region by a factor of $\sim 2$. This difference likely reflects the simplified nature of our model. For instance, Figure~2 of MMA13 includes both hotter and cooler stars than the ones we consider, with correspondingly different rotation properties and spindown efficiencies. Among these objects, late-F dwarf stars, which have a very thin outer convective layer (i.e., a small $I_{\rm e}$) and likely also  much weaker magnetic braking and core--envelope coupling than the Sun, have been proposed as the likely hosts of planetary systems that exhibit near-synchronous rotation \citep[e.g.,][]{MarcyEtal97,Dobbs-DixonEtal04,BarkerOgilvie09}. A well-known example of such a system is $\tau$~Boo, which has a planet with $M_{\rm p} \approx 5.6\, M_{\rm Jup}$ in a synchronous 3.31-day orbit around a dwarf/subgiant F star \citep[e.g.][]{RodlerEtal12}. Model~6 in Table~\ref{tab1} is patterned on the simplified model adopted for this system in Figure~2 of \citet{BarkerOgilvie09}. Figure~\ref{fig2}(d) shows the properties of a hybrid population that contains 5\% late-F hosts, which is roughly the fraction of such KOIs in the MMA13 sample,\footnote{This fraction was estimated using the selection criteria $\log{g/({\rm cm \, s}^{-2})}\ge 4.0$ and $6100\,{\rm K}\le T_{\rm eff} \le 6350\,$K on the surface gravity and effective temperature, respectively, where the lower bound on $T_{\rm eff}$ coincides with the adopted upper limit for solar-type stars in the sample of \citet{Youdin11} and the upper bound is close to the inferred value of $T_{\rm eff}$ for $\tau$~Boo \cite[see][]{RodlerEtal12}. We counted 20 such hosts among the 432 systems in the MMA13 sample that have $P_{\rm orb} \le 10\;$days.} with the rest corresponding to our fiducial (G~dwarf) model. It is seen that, with this admixture, we can qualitatively reproduce the appearance of the $P_{\rm rot} \approx P_{\rm orb}$ systems in Figure~2 of MMA13 as well as in the data presented by \citet{WalkowiczBasri13}. The finding by the latter authors that the synchronicity locus is populated by large planets is consistent with this picture on account of the fact that the strength of the tidal interaction terms in Equations~(\ref{eq:Porb}) and~(\ref{eq:Prot}) is $\propto M_{\rm p}$.  We do not, however, replicate the $P_{\rm rot} \approx 2\, P_{\rm orb}$ locus that these authors discuss: to account for this finding --- assuming that it withstands further observational scrutiny --- would require the inclusion of additional physical processes.\footnote{Note that \citet{Lanza10} identified a similar trend among hot Jupiters associated with hot (F-type) stars with $P_{\rm rot} \lesssim 10\;$days. He proposed an explanation in terms of planet migration through the inner edge of a very long-lived disk that is truncated near the corotation radius by the stellar magnetic field.} The addition of earlier-type hosts also increases the number of nonsynchronous systems with $P_{\rm rot}<10\;$days in our model plot, and this number can be expected to grow as even hotter (more massive) hosts, in which magnetic braking is even less effective, are included.\footnote{Given that late-F stars can be expected to have lower rotation periods, on average, than G dwarfs, we experimented with several initial distributions of $P_{\rm rot}$ for the late-F hosts modeled in Figure~\ref{fig2}(d), with the lowest starting values of $P_{\rm rot}$ ranging from 1 to 10~days. We found that the final planet distribution in the $\log{P_{\rm orb}}-\log{P_{\rm rot}}$ plane was not sensitive to this choice. The results presented in Figure~\ref{fig2}(d) employ the same initial period distribution for F~stars as the one we adopted for G~stars (Figure~\ref{fig1}(a)), which is consistent with the fact that the nominal magnetic braking times for the two model populations used in this calculation are similar.}

Figure~\ref{fig3}(a) shows the final age distribution for a fiducial-model sample of 2000 systems. It is seen that most of the host stars (and particularly those with $P_{\rm rot}>10\;$days) evolve according to the predictions of the gyrochronology method, with older stars characterized by higher values of $P_{\rm rot}$. This validates the use in our Monte Carlo calculations of the age distribution derived by \citet{WalkowiczBasri13} on the basis of this method. This figure also reveals a remarkable property of our model distributions: the entire (oblique) lower edge is defined predominantly by the youngest sampled systems. Although a simple-minded application of gyrochronology would imply that the systems at the lower end of this edge should indeed be the youngest, the unique prediction of our scenario is that systems in the upper part of the lower edge (those with $P_{\rm rot}$ close to $10\;$days) should also be very young and should contrast with systems of comparable rotation periods but longer orbital periods (which would, on average, be measurably older). The relative youth of the systems along the lower edge of the distribution is again the result of magnetic braking, which is continuously acting to slow down the rotation of each of the modeled stars. However, the skewed orientation of this edge, which leads to the predicted behavior, is a consequence of the additional effect of tidally induced planet ingestion. The above prediction clearly cannot be verified through gyrochronology of the host stars, but it could in principle be tested through an alternative age-determination method. One such method, which has already been successfully applied to {\it Kepler}\/ sources, is based on interpreting asteroseismic data on the internal structure of the star in the context of stellar evolution models. This method is particularly robust when continuous observations are conducted over sufficiently long periods to obtain information on individual frequencies \citep[e.g.,][]{MetcalfeEtal10,BatalhaEtal11}, but even in cases where the available photometric information is more limited, one can improve the accuracy with the help of ground-based spectroscopic observations \citep[e.g.,][]{ChaplinEtal14}. Another approach, which can be employed when the host star is a member of a binary system, is to compare the derived ages of the two (presumably coeval) binary companions. This method has also been used in the study of {\it Kepler}\/ stars \citep[e.g.,][]{MetcalfeEtal12,WeisenburgerEtal14} and, in fact, has provided preliminary evidence for the influence of close-in planets on the rotation rates of their host stars \citep[][]{PoppenhaegerWolk13}. An additional noteworthy aspect of our finding that the lower edge of the KOI distribution in the $\log{P_{\rm orb}}-\log{P_{\rm rot}}$ plane is best reproduced by young systems (Figure~\ref{fig3}(a)) characterized by comparatively low values of $Q^\prime_*$ (Figures~\ref{fig1}(b) and~\ref{fig2}(c)) is that this result is consistent with the suggestion \citep[e.g.,][]{Dobbs-DixonEtal04} that $Q^\prime_*$ increases from $\sim 10^5$ to $\sim 10^6$ as a solar-type star ages.

Before closing this section, we comment on the extended planetary distribution function that \citet{Youdin11} derived by extrapolating the discovery efficiencies reported in \citet{HowardEtal12} for $R_{\rm p}>2\, R_\earth$ down to $R_{\rm p} = 0.5\, R_\earth$ (resulting in the addition of 190 KOIs to his sample). We applied our computational scheme to this distribution and obtained a final distribution in the $\log{P_{\rm orb}}-\log{P_{\rm rot}}$ plane that had significantly fewer systems with $P_{\rm rot}<10\;$days than our fiducial model and, correspondingly, a lower edge whose slope was much flatter than that of the MMA13 data. One reason for this discrepancy could be that the Poisson process-based distribution function constructed by \citet{Youdin11} does not accurately describe the properties of multiple planets in any given system, and that in this case our independent drawings for 2- and 3-planet systems do not properly capture the dominant contribution of the more massive planets to the tidal ingestion process. It is, however, also conceivable that even the average number of planets per star is not correctly given by Youdin's extended distribution, and that the relative number of $R_{\rm p} < 2\, R_\earth$ close-in planets is actually smaller than the value ($\gtrsim 1/3$) that this distribution implies \citep[see][]{FressinEtal13}.

\section{Discussion}
\label{sec:discuss}

To determine the effect of the modeled tidal interaction on the adopted initial $P_{\rm orb}$ distribution, we obtained the probability density of the final distribution using Gaussian kernel density estimation. The resulting distribution for our fiducial case (Model~1) is shown by the solid blue curve in Figure~\ref{fig3}(b). It is noteworthy that, even though the initial distribution is a smooth power law in $P_{\rm orb}$, the final distribution manifests a break at $P_{\rm orb} \approx 4\;$days, where it steepens toward lower periods. The location of this break is consistent with the lower bound on the extent of the tidal interaction zone that we inferred from the properties of the final distribution in the $\log{P_{\rm orb}}-\log{P_{\rm rot}}$ plane for this case as well as with the boundary of the void that MMA13 identified in the observed distribution (see Section~\ref{sec:results}). We repeated the calculation for the cases shown in panels (a)--(c) of Figure~\ref{fig2} and determined that the result remains essentially the same when only the stellar spindown parameters are changed (Models~3 and~4), but that the break shifts to $P_{\rm orb} \approx 3\;$days when $Q^\prime_*$ is increased from $10^5$ to $10^6$ (Model~5; dashed red curve in Figure~\ref{fig3}(b)). Interestingly, the location of the latter break is approximately the same as in the data presented by \citet{Youdin11}, where it is fixed largely by the contribution of the $R_{\rm p} > 3\,R_\earth$ planets in his sample. We note in this connection that the values of the power-law index $\beta$ that characterize the asymptotic ``fast'' and ``slow'' regimes for each of the curves in Figure~\ref{fig3}(b) depend on our choice for the upper end of the $P_{\rm orb}$ interval ($P_{\rm orb,max}=10\;$days) and therefore cannot be directly compared with those given in \citet{Youdin11} (where $P_{\rm orb,max}=50\,$ days, and where the inferred values of $\beta$ also depend on the imposed slow/fast separation at $P_{\rm orb}=7\;$days). However, we verified that the presence of the break and its location do not depend on the value of $P_{\rm orb,max}$. These results indicate that the break is a robust feature of the final $P_{\rm orb}$ distribution that reflects the extent of the star--planet tidal interaction zone. The fact that the overall distribution is best represented by a model with 
$Q^\prime_* = 10^6$ even as its edge is better reproduced using $Q^\prime_* = 10^5$ is consistent with the possibility that this  parameter evolves with stellar age (see Section~\ref{sec:results}). Our finding that the same basic model can account for the low-period structure of the KOI distribution in the $\log{P_{\rm orb}}-\log{P_{\rm rot}}$ plane and for the prominent break in the $P_{\rm orb}$ distribution of these objects lends strong support to the tidal interaction interpretation of {\it both}\/ the $P_{\rm orb}$ distribution of close-in planets {\it and} the $P_{\rm rot}$ distribution of their host stars.

\acknowledgments
We are grateful to Tsevi Mazeh for bringing the issue considered in this paper to our attention. We thank him and Dan Fabrycky for many enlightening discussions, and Rich Kron, Yoram Lithwick, Titos Matsakos, and Fred Rasio for helpful input. We also acknowledge the anonymous referee's useful suggestions for improving the manuscript. This research was supported in part by NASA ATP grant NNX13AH56G.

\end{document}